\documentclass{menu97}
\usepackage{epsfig}
\begin{document}
\setlength{\baselineskip}{2.6ex}

\title{$\pi Y$ Phenomenology from $\bar KN$ Scattering\thanks{Research 
supported partially by the E.E.C. Human Capital and Mobility Program under 
contract No. CHRX--CT92--0026.}}

\author{Paolo M. Gensini, Rafael Hurtado\thanks{On leave of absence from 
Centro Internacional de F\'\i{sica}, Bogot\'a, Colombia, under a grant by 
Colciencias.}\\{\em Dip. Fisica, Univ. Perugia, and Sez. I.N.F.N., Perugia, 
Italy}\\and\\Galileo Violini\\{\em Dip. Fisica, Univ. Calabria, and Gr. Coll. 
I.N.F.N., Cosenza, Italy\\and\\Univ. de El Salvador, San Salvador}}

\maketitle

\begin{abstract}
The dispersive determination of $\pi Y\Sigma$ coupling constants ($Y= \Lambda, 
\Sigma$) and the implications for low-energy $\bar KN$ analysis are discussed.
\end{abstract}

\section*{1. Introduction.}

There are two reasons of interest for determining the $\pi Y\Sigma$ coupling 
constants: the existence of definite predictions for the $PBB$ couplings by 
flavour SU(3) symmetry \cite{1}, and the forthcoming opening of 
DA$\Phi$NE \cite{2} that will create conditions which should improve the 
experimental knowledge of the S = -- 1 meson--nucleon interaction.

The current knowledge of $PBB$ couplings is that $G_{\pi NN}$ is known within 
a few percent from $\pi N$, $NN$, and $\bar NN$ scattering \cite{3}. The two 
$\bar KNY$ couplings are known with less, but still reasonable, accuracy from 
$KN$ dispersion relations \cite{4}. A few papers can also be found 
dealing with the determination of $\pi\Lambda\Sigma$ \cite{5,6,7,9,10} and 
$\pi\Sigma\Sigma$ \cite{5,7,8,9} couplings, whereas for the $\eta NN$ 
coupling real attempts at its determination have been even scarcer.

Processes like $\pi\Lambda$ and $\pi\Sigma$ elastic scattering, or 
$\pi\Lambda\to\pi\Sigma$, are not accessible to experiment: however, 
parametrizations for their amplitudes are provided as a byproduct of the 
coupled--channel K-- or M--matrix formalism employed to analyse $\bar KN$ 
interactions. In its simplest form, this formalism considers two--body 
channels only, and, when used in the kinematical region between the $K^-p$ 
threshold and an energy just above the $\Lambda$(1520) D--wave resonance, 
it has allowed to determine, by analytic continuation, the $\bar KN$ 
elastic amplitudes in the unphysical regions between the $\pi\Lambda$ 
(for the I = 1 channel) or $\pi\Sigma$ (for the I = 0 one) thresholds 
and the elastic one, describing as well the amplitudes for all the 
other channels in a framework that guarantees unitarity. Of course, 
$\bar KN$ reactions can only provide information on the I = 0, 1 
channels, so that the I = 2 $\pi\Sigma$ amplitude remains unknown.

We use the amplitudes obtained in this way to calculate the $\pi Y\Sigma$ 
couplings by conventional dispersion relations \cite{11}  applied to 
three reactions, namely elastic $\pi\Lambda$ and $\pi\Sigma$ scattering 
and $\pi\Lambda\to\pi\Sigma$: for the first two cases dispersive 
calculations may already be found in the literature, whereas the 
third amplitude is discussed here for the first time in this context. 
In a second paper presented at this Symposium, more sophisticated 
techniques, also based on analyticity, will be reported \cite{12}.

Our purposes are, first, to update the results of refs. [5,6], making use of 
updated inputs in various energy ranges of the dispersive integrals; second, 
to analyse the stability of the results against variations of the energy 
$\nu$ at 
which the DRs are evaluated, whereas all previous works used DRs at just a 
single energy: this, together with the stability against the use of 
different dispersive techniques, constitutes a test of the goodness of the 
K-- or M--matrix elements for the physically unfeasible $\pi Y$ reactions. 
Last but not least, we wish to analyse the behaviour of the contributions to 
the couplings from each partial wave, in order to understand whether suitable 
experiments at DA$\Phi$NE might improve our knowledge on the $\pi Y$ 
amplitudes. It has to be remembered that the experiments used in the past to 
determine the K-- or M--matrix elements were performed with statistics much 
lower than those attainable at DA$\Phi$NE, and that this machine will be 
able to explore in detail the very-low energy region around the $\bar K^0n$ 
charge-exchange threshold, where energy--dependent effects due to the $K^-p$ 
-- $\bar K^0n$ mass difference are expected to show up. This strategy has 
proved to be useful, since we shall show that the values of the coupling 
constants exhibit a variation with energy in the region studied, and shall 
also point to the partial wave(s) presumably responsible for this behaviour.

The following part of this paper will describe the methodology of our work, 
including the selection of inputs made to evaluate the DRs, the third part 
will discuss each one of the processes (namely elastic $\pi\Lambda$ and 
$\pi\Sigma$, and $\pi\Lambda\to\pi\Sigma$), and, finally, we shall present 
some comments on our results.

\section*{2. General formalism.}

Standard DRs allow to evaluate the coupling constants as sum of two terms, 
proportional, respectively, to the real part of the amplitude evaluated at a 
suitable energy $\nu$ and to the dispersive integral, which ranges from the 
lowest threshold (with the same quantum numbers as the process considered) 
to infinity. It is convenient to separate the contributions to the integrals 
into three energy regions: low, intermediate and high. The choices of the 
DR, of the $\nu$ value, and of the amplitudes to which to apply the DR, 
depend on the quality of the inputs at different energies. We use the 
fastest--converging dispersion relations, in order to reduce the uncertainty 
coming from the high--energy region, since in our case the most reliable 
inputs should be those provided by the low--energy K-- or M--matrix 
analyses. The integrals in the energy range above those considered by 
these latter have been estimated using resonance couplings at intermediate 
energies and a Regge--pole model at the highest ones, and matching the 
two through two--component duality \cite{13}. In the absence of a point 
at which a subtraction could be done without extra theoretical inputs, 
such as e.g. the Adler consistency condition \cite{7,8,9}, the 
best alternative turns out to be the use of unsubtracted DRs.

The criteria just expressed for the DR lead to choose amplitudes which 
have the fastest--decreasing behaviour as $\omega\to\infty$: in terms of 
the invariant amplitudes $A(\omega,t)$, $B(\omega,t)$ and $C(\omega,t)$ 
(with $C(\omega,t) = A(\omega,t) + \omega B(\omega,t)$ being the 
amplitude whose imaginary part in the forward direction is given by the 
optical theorem as $k\sigma_{tot}(k)/{4\pi}$), we use $B(\omega,0)/\omega$ 
for the elastic $\pi\Lambda$ and $\pi\Sigma$ scattering, and 
$C(\omega,0)/\omega$ for the inelastic process $\pi\Lambda\to\pi\Sigma$.

$A$ and $C$ are dominated by the S--wave, while $B$ by the P--waves: 
since some of the K--matrix analyses \cite{14,15} are purely S--wave, 
and the existence of the $\Sigma$(1385) requires at least inclusion of 
the P$_{13}$ one as well, a calculation using such parametrizations 
would be useless for $B$, whereas for $A$ and $C$ would lead to 
determine an ``effective coupling'' for both the $\Sigma$ and the 
$\Sigma$(1385).

Due to the absence of information on the I = 2 $\pi\Sigma$ amplitude, in the 
elastic $\pi\Sigma$ DR we are forced to use the crossing-even combination of 
isospin amplitudes $2B_1(s,t,u)-B_0(s,t,u)$, which is the only one free 
of the I = 2 part both in the $s$ and $u$ channel: an identical choice 
was made in the earlier applications of DR to this process \cite{5,7}.

For the low--energy region there exist two {\em published} K-- or M--matrix 
parametrizations 
extending down to the $\bar KN$ threshold and including S--, P-- and D--waves 
\cite{16,17}. The second, which includes also F--waves, is a fit amalgamating 
different analyses, but it does not show the $\Lambda$(1405) and 
$\Sigma$(1385) features below the $\bar KN$ threshold: this leaves 
Kim's parametrization \cite{16} as the only viable choice for our 
calculations. Kim's analysis covers the region from threshold to 550 MeV/c, 
and expresses the T matrix as 
\begin{equation}
T_l={q^{2l} \over M-iq^{2l+1}},\ \ \ \ {\rm with} \ \ \ \ M(q^2)=M(q_0^2)+{1 
\over 2} \delta_{ij} r^{1-2l} C_l(q^2-q^2_0) ;
\label{LABELLING} \end{equation}
here $q$ is the momentum matrix in the c.m. frame and $q_0$ its value at the 
$\bar KN$ threshold, the M--matrix is expressed in the effective--range 
approximation, $l$ is the angular momentum, $r$ the diagonal effective--range 
matrix and $C_l$ an appropriate coefficient. This analysis gives a good 
description of S and P$_{13}$ waves, whose energy dependence includes the 
first and second coefficient of the effective range expansion, whereas that 
of P$_{01}$, P$_{11}$ and P$_{03}$ waves is poor, since their energy 
dependences include only the first coefficient: the development in powers 
of $q^2$ being not the same for all waves, the fact that this might 
be the source of some internal inconsistency has been noticed in 
ref. [18].

In the intermediate--energy region (550 MeV/c to 2,000 MeV/c), we saturate 
the integrals by resonances in a narrow--width approximation, using the 
parametres of ref. [19]: we have preferred this to the PDG compilation in 
order to have a self--consistent analysis in the whole region. The 
resonance contribution is not unambiguous \cite{6}, and it can also be 
noticed that the $\Lambda$ and $\Sigma$ resonances actually observed are 
about one--half the number expected on the basis of SU(6) 
and of the known $N$'s and 
$\Delta$'s: however, the total contribution from this region is small, so 
that these uncertainties are not very relevant in our calculation.

In the intermediate and high--energy regions, the background contributions 
for $\pi\Lambda$ and $\pi\Sigma$ elastic scattering are calculated from 
Pomeron exchange, assuming a multi--component model to account for the 
deviation of the asymptotic ratios $\sigma_{tot}^{K^+p} / 
\sigma_{tot}^{\pi^+p}$ and $\sigma_{tot}^{pp}/\sigma_{tot}^{\pi^+p}$ from 
1 and 1.5 respectively. The only attempt to evaluate Regge--pole couplings 
for $\pi Y$ scattering was made only for the Pomeron coupling to 
$\pi\Lambda$ using FESRs \cite{20}: for a more complete description of 
the asymptotic region we extend the Regge--pole fit by Donnachie and 
Landshoff \cite{21} to all $\pi N$, $NN$ and $\bar NN$ data above $s$ = 6 
GeV$^2$, using for the Reggeon couplings flavour SU(3) symmetry and the 
Okubo--Zweig rule, and assuming their parametrization \cite{21} to extend 
to energies much lower than those where it was fit to the data: we 
thus obtain $\sigma_{tot}^{\pi\Lambda(\Sigma)}$ = 12.42 mb $\cdot 
(s/s_0)^{0.0808}$ + 25.13 mb (17.25 mb) $\cdot (s/s_0)^{-0.525}$ (with 
$s_0^{1/2}$ = 1 GeV). Working with $B$, rather than with $C$ as in the 
fit by ref. [21], we then resort to the approximate relation $B \simeq 
C/\omega$, valid only for the dominant asymptotic terms in an optical model.

\section*{3. Applications.}

The process $\pi\Lambda\to\pi\Lambda$ is pure I = 1, with just one Born term 
in the DR, whose pole at $t$ = 0 lies at $\omega_\Sigma$ = 
71.35 MeV, well below 
the threshold $\omega_{th}=m_\pi$. The contributions to the unsubtracted DR 
for $B/\omega$ can be rated according to their magnitude, being small those 
from the low-energy S--wave, Regge poles, the Pomeron and 
intermediate--energy resonances, leaving as the most important ones those 
from the low--energy P$_{11}$ and P$_{13}$ waves.

It is important to notice that the main contribution to the coupling constant 
comes from the poorly known P$_{11}$ wave (Fig. 1.a): indeed, the huge 
variation with energy of $G^2_{\pi\Lambda\Sigma}$ reflects the similar 
behaviour of this contribution. Although this kind of calculation can only 
be performed using parametrizations which include P--waves, this result casts 
doubts on the reliability of Kim's one \cite{5}. It should also be noted 
that, if the contribution from the P$_{11}$ wave is subtracted, the sum of 
the remaining contributions to $G_{\pi\Lambda\Sigma}^2$ remains almost constant 
in a wide energy interval. Table 1 shows the values obtained in different 
works; our results are presented both with all contributions (a), and 
having subtracted the P$_{11}$ one (b).

For the process $\pi\Sigma\to\pi\Sigma$ there are two Born terms, one for 
the $\Lambda$ and one for the $\Sigma$. The DR for this process exhibit a 
peculiarity with respect to other elastic meson--nucleon DRs, since the 
$\Lambda$ pole falls on the cut between the $\pi\Lambda$ and $\pi\Sigma$ 
thresholds. Thus we can not expect the M--matrix to describe correctly the 
real part of the amplitude close to the $\Lambda$ pole where, however, 
the imaginary part of the amplitude is not affected by this fact. At 
energies away from the position of the $\Lambda$ pole one expects a smooth 
behaviour of both the real and imaginary parts, so that the DR can be safely 
evaluated there, except near the cutoff just above 1,600 MeV, where effects 
of the poor matching with the resonance saturation region appear. If we 
use as additional input in the $\pi\Sigma$ DR the value of 
$G^2_{\pi\Lambda\Sigma}$ obtained from the $\pi\Lambda$ one, we find again 
a strong variation with energy of $G^2_{\pi\Sigma\Sigma}$ as well. 
Also this variation can be attributed to P--waves effects, since 
subtraction of the contributions from P$_{01}$ and P$_{11}$ waves eliminates 
this structure (Fig. 1.b); again Table 1 presents the values obtained in 
different works for $G^2_{\pi\Sigma\Sigma}/4\pi$, together with our 
results both with and without the J = $\frac{1}{2}$ P--waves.

{\small
\begin{center}
\centerline{\bf Table 1}
$G^2_{\pi\Lambda\Sigma}/4\pi$ from $\pi\Lambda\to\pi\Lambda$ and 
$G^2_{\pi\Sigma\Sigma}/4\pi$ from $\pi\Sigma\to\pi\Sigma$
\begin{tabular}{||r|r|l||} \hline 
$\bf{G^2_{\pi\Lambda\Sigma}/4\pi}$ & $\bf{G^2_{\pi\Sigma\Sigma}/4\pi}$ & 
\bf{Method and Refs.}  \\ \hline 
$21.5 \pm 7$ & $11.4 \pm 5$ & DR \cite{5} \\ \hline
$16.5 \to 19.1$ & & DR \cite{6}\\ \hline 
$20.9 \pm 6.7$ & $11.4 \pm 5.5$ & DR + Adler C.C.  \cite{7}\\ \hline 
$12.9 \pm 0.8$ & $12.5 \pm 2$ & FCDR + Regge \cite{8}\\ \hline
$12 \pm 1$ & $13 \pm 1$ & DR + Adler C.C.  \cite{9}\\ \hline 
$17.5$ at $w^{\bar KN}_{th}$ & $11.2$ at $w^{\bar KN}_{th}$ & DR (a)\\ \hline 
$5.0$ at $w^{\bar KN}_{th}$ & $3.9$ at $w^{\bar KN}_{th}$ & DR (b)\\ \hline 
\end{tabular}
\end{center}}

\begin{center}
\epsfig{figure=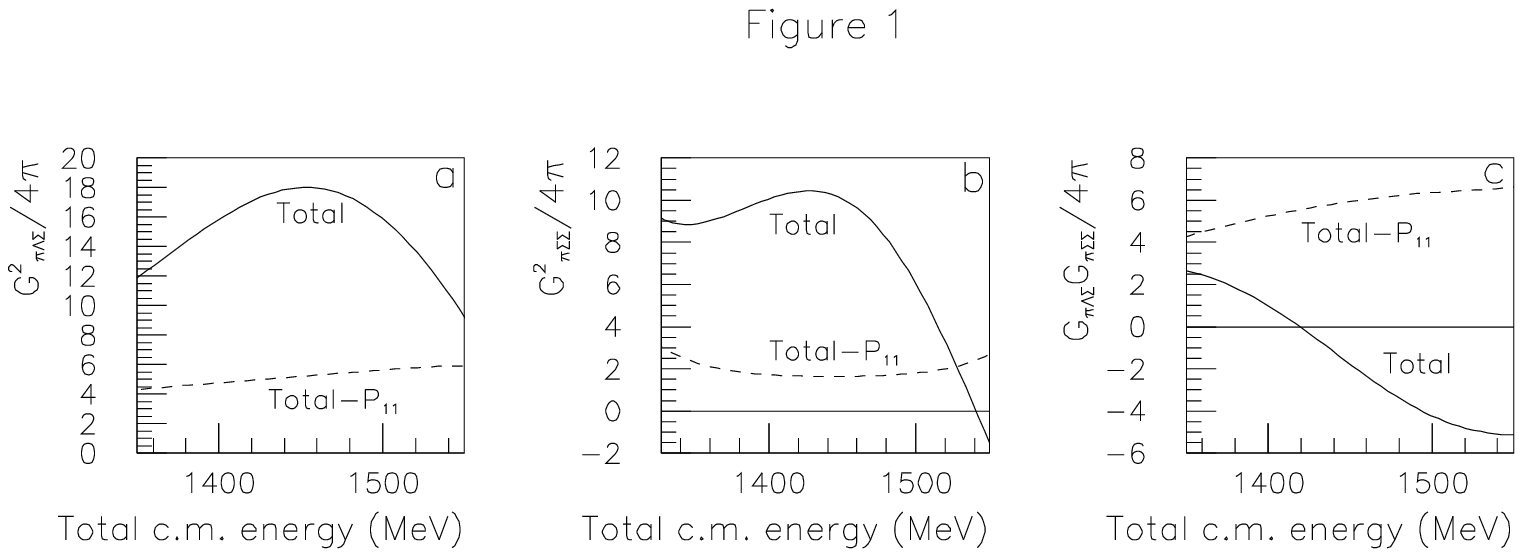,width=16cm,height=6cm}
\end{center}

For the reaction $\pi\Lambda\to\pi\Sigma$, which is a pure I = 1 process, 
and whose amplitudes have the same analytical structure as the elastic 
$\pi\Lambda$ ones, we use the crossing-even superconvergent amplitude 
$C/\omega$: of course in this case the Born term is proportional to the 
product of the two $\pi Y\Sigma$ coupling constants.

The results which include the contributions from all partial waves are not 
consistent with the previous two calculations (Fig. 1.c): the product 
$G_{\pi\Lambda\Sigma}G_{\pi\Sigma\Sigma}$ exhibits a variation from 
positive to negative values in the region around the $\bar KN$ 
threshold. Looking at the contributions from the different partial waves, 
one again sees that the contribution responsible for the change of sign 
is that from P$_{11}$, and that the dominant ones are S-- and P--waves, 
whereas intermediate--energy resonances and Regge--poles are negligible.

There is a clear inconsistency between a near--zero value 
for the product at the 
$\bar KN$ threshold and the values of the coupling constants obtained from 
elastic $\pi Y$ scattering at the same c.m. energy. However, the value 
corresponding to subtraction of the P$_{11}$ wave, 
$G_{\pi\Lambda\Sigma}G_{\pi\Sigma\Sigma}/4\pi\simeq$ 5.0 
at $w^{\bar KN}_{th}$, 
is very close to the values obtained for both $G_{\pi\Lambda\Sigma}^2/4\pi$ 
and $G_{\pi\Sigma\Sigma}^2/4\pi$ from elastic $\pi Y$ DRs ignoring the 
contributions from P$_{01}$ and P$_{11}$ waves. This seems to suggests that 
J = $\frac{1}{2}$ P--waves are poorly determined in Kim's analysis, or at 
least that their $\pi Y$ matrix elements are. Moreover, it should be noted 
that even if $C$ is commonly assumed to be dominated by the 
S--wave, the importance of the P--waves contributions in the present 
case should lead to question any result based on a pure--S--wave 
parametrization.

\section*{4. Conclusions.}

DR calculations of $\pi Y\Sigma$ couplings indicate that 
current $\bar KN$ low--energy analyses do not adequately determine all 
P--waves for the coupled $\pi Y$ channels. This does not allow to determine 
the $G_{\pi Y\Sigma}$ coupling constants with the accuracy claimed by 
previous calculations. If the J = $\frac{1}{2}$ P--wave contributions were 
much overestimated, then it would still be possible for these couplings to 
agree with each other and with SU(3)--symmetry predictions. This work 
suggests also to explore whether the $\bar KN$ matrix elements suffer from 
the same problem: 
in fact, an analogous study for $\bar KN$ scattering \cite{18} found, 
working with once--subtracted DRs for the $C$ amplitudes, an energy 
dependence of the coupling $G_{KN\Lambda}^2$ with a slope 
$d(G_{KN\Lambda}^2/4\pi)/d\nu\simeq$ 30 GeV$^{-2}$ at the $\bar KN$ 
threshold, an effect 
which perhaps, on the basis of what said above, should have been attributed 
to the inadequacy of Kim's ``small'' P$_{01}$ and P$_{11}$ waves.

The above results have shown that the low--energy S = -- 1 meson--nucleon 
sector is less well known than currently believed, and that it deserves 
more accurate experiments and analyses if one wants to reach for its 
knowledge a level comparable to that of the $\pi N$ one. DA$\Phi$NE is the 
right machine to achieve this goal \cite{22}.

\bibliographystyle{unsrt}

\end{document}